\documentclass[11pt,twoside]{article}

\usepackage{asp2004}
\usepackage{epsf}
\usepackage{psfig}
\usepackage{lscape}
\usepackage{graphicx}

\begin{document}
\title{Optical \& Infrared Observations of Stellar Mass Loss in Globular Cluster Red Giants}    
\author{Iain McDonald$^1$, Jacco Th.~van Loon$^1$}   
\affil{$^1$Astrophysics Group, Lennard Jones Laboratories, EPSAM, Keele University, Staffordshire, ST5 5BG, UK}    

\begin{abstract} 
We are examining mass loss from globular clusters giant stars, focussing on metallicity dependance. We present three sets of observations: TIMMI-2 mid-IR spectra of 47 Tuc, UVES high-resolution optical spectra of several clusters, and an infrared atlas of $\omega$ Cen using the Spitzer Space Telescope.
\end{abstract}


\section{Introduction}

Mass loss from giants is of great importance to stellar evolution --- low-mass stars lose up to 0.2 M$_{\odot}$ on the giant branch. Metal-poor globulars giants can probe metallicity dependances of this mass loss, exposing the Universe's chemical enrichment history, and revealing how low-metallicity stars can produce large quantities of dust (\emph{e.g.} \citeauthor{BWvL+06}~\citeyear{BWvL+06}).

\section{Mid-Infrared Spectroscopy in 47 Tucanae}

Eight infrared-excessive giant branch stars in 47 Tuc were observed in the mid-infrared with the ESO La Silla 3.6m and TIMMI-2 spectrograph. Clear silicate emission is present in V1, corresponding to a mass loss rate of 10$^{-6}$ M$_{\odot}$ yr$^{-1}$ (estimated using the {\sc dusty} code --- \cite{INE99}). This work was published as \cite{vLMO+06}.

\section{Optical Spectroscopy of Cluster Giants}

\begin{figure}[!ht]
\resizebox{\hsize}{!}{\includegraphics{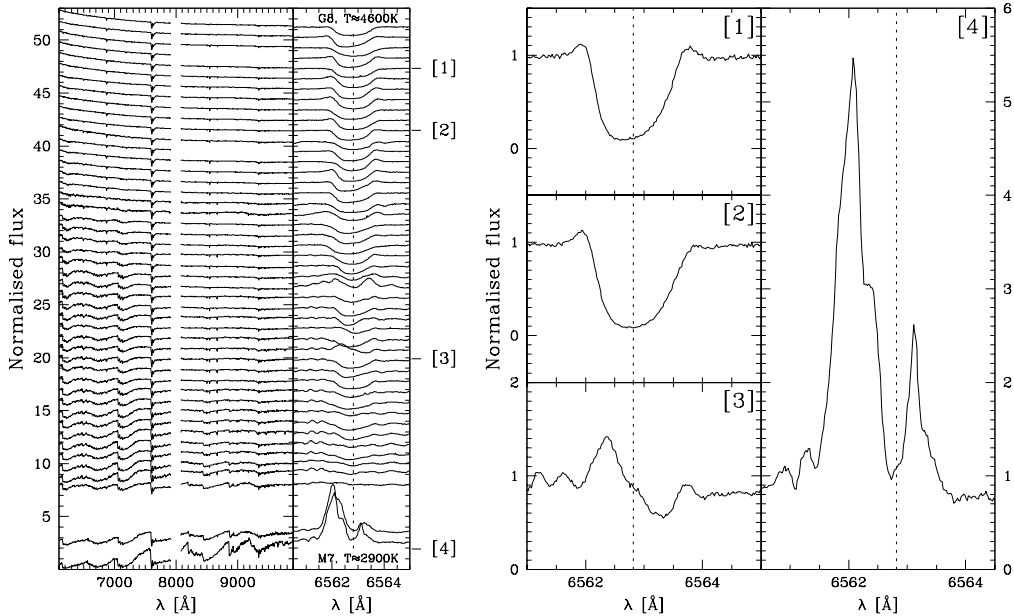}}
\caption[]{[Left] UVES spectra, sorted by spectral type and normalised to their respective 7000 \AA\ fluxes. The spectra have been smoothed for clarity. The central inset details the H$\alpha$ line. [Right] Example profiles from the inset, showing examples of core asymmetries, emission and emission asymmetries and pulsation-shocked emission.}
\label{SpectraFig}
\end{figure}

Optical VLT/UVES spectra were taken of 47 giant stars in six clusters, shown in Fig.~1, in order to analyse differences between IR-excessive and normal stars at a range of metallicities. At $R >$ 100,000, these represent some of the highest resolution data ever taken of globular cluster giants.

The H$\alpha$ and mid-IR Ca II triplet lines exhibit evidence for mass loss: core shifts of several km s$^{-1}$ exist in many stars, as does substantial emission. We are in the process of qualitatively analysing the H$\alpha$ lines using the {\sc sei} code \citep{LCSP87} and various other computational methods. We will then perform an abundance analysis and attempt to model the chromospheric emission.

\section{Spitzer Space Telescope Infrared Atlas of $\omega$ Centauri}

Spitzer imaging of $\omega$ Cen totalling 18.4 hours was taken in six bands from 3.6--70 $\mu$m. We will produce the first multi-wavelength atlas of $\omega$ Cen in the infrared. We can then use this to find dust-enshrouded objects in the cluster, and measure their mass loss rates and other characteristics.

\acknowledgements 
We are thankful to JPL/NASA and ESO for the use of their facilities. Iain McDonald is supported by a PPARC studentship.


\end{document}